# Ranking and mapping of universities and research-focused institutions worldwide based on highly-cited papers: A visualization of results from multi-level models

Lutz Bornmann[1], Moritz Stefaner[2], Felix de Moya Anegón[3], & Rüdiger Mutz[4]

[1] Division for Science and Innovation Studies, Administrative Headquarters of the Max Planck Society, Munich, Germany

[2] Eickedorfer Damm 35, 28865 Lilienthal

[3] CSIC/CCHS/IPP, Albasanz 26 Madrid, Spain

[4] Professorship for Social Psychology and Research on Higher Education, ETH Zurich, Zurich, Switzerland

**Abstract**

Purpose: The web application presented in this paper allows for an analysis to reveal centres of excellence in different fields worldwide using publication and citation data. Only specific aspects of institutional performance are taken into account and other aspects such as teaching performance or societal impact of research are not considered.

Design/methodology/approach: Based on data gathered from Scopus, field-specific excellence can be identified in institutions where highly-cited papers have been frequently published.

Findings: The URL of the web application is as follows: http://www.excellencemapping.net. The web application combines both a list of institutions ordered by different indicator values and a map with circles visualizing indicator values for geocoded institutions.

Originality/value: Compared to the mapping and ranking approaches introduced hitherto, our underlying statistics (multi-level models) are analytically oriented by allowing (1) the estimation of values for the number of excellent papers for an institution which are statistically more appropriate than the observed values; (2) the calculation of confidence intervals as measures of accuracy for the institutional citation impact; (3) the comparison of a single institution with an "average" institution in a subject area, and (4) the direct comparison of at least two institutions.

# 1    Introduction

There is a growing interest in national and international comparisons of research organizations and urban areas in terms of scientific output and impact. A sign of this trend is the continuous publication of university rankings, both inside but also outside the scientific environment (Shin et al., 2011). For example, the SCImago Research Group (University of Granada, Spain) (2012) publishes annually an international ranking of more than 3,000 research institutions and organizations. The reports show indicator values (e.g. publication output, relative citation rate, or excellence rate) based on publication and citation data from Scopus (Elsevier) for larger research-focused institutions (see here SCImago Reseach Group, 2011). The Leiden Ranking (http://www.leidenranking.com/) measures the scientific performance of 500 major universities worldwide (Leydesdorff and Bornmann, 2012b, Waltman et al., 2012). In both rankings, institutions and organizations are listed one after another and can be directly compared in terms of different indicators for productivity, research impact, specialization, and collaboration. Other academic rankings basically follow the same approach but use more or other indicators (see overviews in Kroth and Daniel, 2008, Buela-Casal et al., 2007, Shin et al., 2011).

According to Tijssen et al. (2002), Tijssen and van Leeuwen (2006) as well as Waltman et al. (2012), excellent or highly cited papers are those among the 10% most-cited papers in a field (papers in or greater than the $90^{th}$ percentile, referred to in the following as class 10% papers). According to Waltman et al. (2012) the number of class 10% papers is "the most important impact indicator" for the ranking of universities by research performance. Some different approaches have been published recently which visualize the number of excellent papers for locations of research on Google maps instead of the presentation of the numbers in long ranking lists. Frenken et al. (2009) suggest grouping such approaches to mapping the geography of science under the heading "spatial scientometrics." The visualizations identify regions of excellent research and allow the comparison of excellent output in regions worldwide. Leydesdorff and Persson (2010) explore the use of Google Maps and Google Earth for generating spatial maps of science. Bornmann et al. (2011a) present methods to map centres of scientific excellence around the world. By colorizing cities worldwide according to the output of excellent papers, their maps provide visualizations where cities with a high (or low) output of these papers can be found. Bornmann and Waltman (2011) follow their approach in general, but change the focus from mapping of single cities to a more "sliding" visualization of broader regions. The maps generated by Bornmann et al. (2011a) and Bornmann and Waltman (2011) for different scientific fields point out a spatial concentration of excellent research which might possibly be explained by the fact that more competitors (here: prolific scientists) working within the same region produce better results (Bornmann et al., 2011a).

The most recent stage of the developments in "spatial scientometrics" is the approach of Bornmann and Leydesdorff (2011). They consider it a disadvantage that only the output of excellent papers is used in the other approaches. If a city has a very high output in general (that means not only of excellent papers) one can expect a high number of excellent papers proportionally (Bornmann et al., 2011b, Bornmann et al., 2011c). Therefore, the observed output should be compared with an expected output. For example, if authors at a university have published 10,000 papers, one would expect for statistical reasons that approximately one thousand (that is, 10%) would also belong to the class 10% papers. An observed number of 700 highly-cited papers for this university may seem a large number compared to other universities, but it turns out to be smaller than one would expect in this case. The $z$ test for comparing a population proportion with a sample proportion (Sheskin, 2007) can be used for evaluating the degree to which an observed number of class 10% papers differs from the expected value (Bornmann et al., 2012).

In this paper, we introduce a new web application which is linked to both spatial visualization approaches as well as academic ranking lists published hitherto. It tries to capture the advantages of both approaches for presenting the research performance of locations; lists and maps are visualized and intertwined. In ranking lists, one can immediately see the best and worst institutions (institutions in the first and last positions). However, it is difficult to directly compare institutions holding very different positions or which are located in a single region (e.g. Europe). In contrast, performance indicators visualized on Google maps allow the focus on institutions in certain regions, but it is difficult to identify the best and worst locations worldwide. The web application introduced here visualizes institutional performance within specific subject areas as ranking lists and on custom tile-based maps. In contrast to many other university rankings which present the results across all fields of science (e.g. the Leiden Ranking), the lists and maps shown in the web application are differentiated for subject categories.

Based on the developments of Bornmann and Leydesdorff (2011) we compare in this study the number of observed with the number of expected papers belonging to class 10% within their field category for universities and research-focused institutions (referred to as institutions in the following) around the world. Bornmann and Leydesdorff (2011) conduct a single test for each city worldwide to analyze the statistical significance of the difference between observed and expected numbers (Leydesdorff and Bornmann, 2012b, Bornmann et al., 2012). In this study, a multilevel logistic regression analysis is calculated to analyze the differences between both numbers for all the institutions within one model: For each institution, the difference between its performance and the average over all institutions in a field is tested. The model allows the calculation of shrinkage estimates and corresponding standard errors which are more precise than raw probabilities (empirical Bayes estimates) and their standard errors, especially if the information for an institution is sparse (e.g. if its publication output is low). The estimated standard errors and corresponding confidence interval of the regression model takes design effects into account which are on average higher than the corresponding standard errors and confidence intervals obtained in a sampling procedure which does not consider any clusters, i.e. institutions (Hox, 2010). Additionally, multilevel models provide a very easy way to compare institutions, that is, whether they differ statistically significantly in their probabilities of having published excellent papers.

## 2 Methods

### 2.1 Data sets

The study is based on Scopus data (Elsevier) which has been collected for the SCImago Institutions Ranking (http://www.scimagoir.com/). To obtain reliable data in terms of geo-coordinates (see Bornmann et al., 2011a) and the number of excellent papers (Waltman et al., 2012), we consider in the study only those institutions that have published at least 500 articles, reviews and conference papers in the period 2005 to 2009 in a certain Scopus subject area. Institutions with fewer than 500 papers in a category are not considered. Furthermore, only subject categories offered at least 50 institutions are included in the web application (e.g. Arts and Humanities is not included). We use this threshold to have a considerable number of institutions for a worldwide comparison. The full counting method was used (Vinkler, 2010) to attribute papers from the Scopus data base to institutions: if an institution appears in the affiliation field of a paper, it is attributed to this institution (with a weight of 1). According to the results obtained by Waltman et al. (2012) the overall correlation between a university ranking based on the full counting and fractional counting method is very high (r = .97). The fractional counting method gives less weight (<1) to collaborative than to non-collaborative papers (= 1). Table 1 shows the number of institutions which are considered as data sets for the 17 subject areas in this study. Out of the 27 available subject areas in Scopus, only those are selected for the study which include at least 50 institutions worldwide.

When evaluating the citation impact of publications there is the possibility of including or excluding authors' self-citations (Bornmann et al., in press-a). Studies have reported different percentages of self-citations: For example, in a study on researchers in Norway, Aksnes (2003) found 36%; Snyder and Bonzi (1998) showed that the percentage of self-citations in natural sciences is 15%; higher than in social sciences (6%) and arts & humanities (6%). In this study, we included self-citations for two reasons: (i) It is not expected that the percentage of self-citations will differ significantly among the different authors at the institutions. The percentage of self-citations will vary among the authors (and the publications), but in most cases it is not expected that institutions conducting research in similar areas have very different self-citation percentages. (ii) Following Glänzel et al. (2006), self-citations are usually an important feature of the science communication and publication process: "A self-citation indicates the use of own results in a new publication. Authors do this quite frequently to build upon own results, to limit the length of an article by referring to already published methodology, or simply to make own background material published in 'grey' literature visible" (p. 265).

## 2.2 Percentile calculation

To identify the class 10% papers within a subject area, the citations $X_i$ (citation window: from publication until the end of 2011) that were received by the $i$th papers within $n$ papers published in a given subject area (and publication year as well as a given document type) were gathered. Then the papers were ranked in increasing order

$X_1 \leq X_2 \leq \ldots \leq X_n$,

where $X_1$ and $X_n$ denote the number of citations received respectively by the least and most cited paper. Where citation counts are equal, the SJR2 (Guerrero-Bote and de Moya-Anegon, 2012) of the journal which has published the papers is used as a second sort key (from highest to lowest). This journal metric takes into account not only the prestige of the citing scientific publication but also its closeness to the cited journal. Finally, in each field (publication year and document type), each individual publication was assigned a percentile rank based on this distribution. If, for example, a single paper within a subject area had 50 citations, and this citation count was equal to or greater than the citation counts of 90% of all papers in the subject area, then the percentile rank of this paper would be 90. The paper would be in the 90th percentile and would belong to the class 10% papers within the subject area. There are different approaches available for calculating percentile-based indicators (see an overview in Waltman and Schreiber, 2013, Bornmann et al., 2013). The approach used for the SCImago Institutions Ranking is comparable to the approach proposed by Rousseau (2012).

In Table 1, the mean percentage of highly-cited papers for the institutions included in this study is the mean over the percentages of class 10% papers for the single institutions within one subject area. For example, Physics and Astronomy consists of 650 different institutions with a mean proportion of excellent papers of 0.14. Three reasons can be cited for the fact that the mean average for Physics and Astronomy (as well as all other subject areas in the table) is higher than 10%: (i) Ties in citation data lead to a higher number of class 10% papers (Leydesdorff et al., 2011, Waltman and Schreiber, 2013). (ii) The highly-selected set of institutions considered here (institutions with at least 500 publications) has published more class 10% papers than institutions not considered. (iii) "First, collaborative publications are counted multiple times in the full counting method, and second, collaborative publications tend to be cited more frequently than non-collaborative publications" (Waltman et al., 2012, p. 2427).

### 2.3 Statistical model

The choice of the statistical procedure to analyze the data depends strongly on the scale of the dependent variable (e.g. ordinal, continuous). In our case the dependent variable is dichotomous: a paper published by an author located in an institution belongs to the class 10% papers or not. The relative frequency of the papers in the class 10% for an institution is an estimate of its probability of class 10% papers. The simplest way for the statistical analysis is to report these probabilities of class 10% papers for each institution (see Waltman et al., 2012). However, this procedure is statistically not appropriate and leads to incorrect solutions, because the hierarchical structure of the data (papers, level 1, are nested within institutions, level 2) is not taken into account. We can assume that papers, published by authors within one institution, are somewhat more homogeneous regarding their probability of being class 10% papers than papers published by authors located in different institutions. The homogeneity reduces the effective sample size and increases the standard errors.

We prefer a multilevel logistic regression (MLR) intercept-only model for binary outcomes, which properly estimates the standard errors. Not only the standard errors of the regression parameter, but also the size of the standard error of the estimated class 10% probabilities (might) differ from those of a one level model with consequences for the statistical comparison of institutions. Another great advantage of multilevel modelling is that the statistical results can be summarized with a small set of parameters. For instance, one parameter allows to test statistically, whether the institutional performances vary only by random (i.e. as random samples of the same population) or systematically. Only in the case of systematic differences between institutions, comparisons are reasonable.

In MLR, papers are clustered within universities, whereas j (j = 1, …, N) denotes the level-2 units ("institutions") and i (i = 1, …, $n_j$) the level-1 units ("papers"). Due to the fact that the dependent variable $x_{ji}$ is dichotomous (1 = paper i belongs to the class 10% publications, 0 = paper i does not belong to the class 10% publications), ordinary multilevel models for continuous data are not appropriate. Therefore, so-called generalized linear mixed models are favoured, especially, the multilevel logistic model for binary data, which comprises three components (Hox, 2010):

1. The probability distribution for $p_{ji}$ (= $Pr(x_{ji}=1)$) is a Bernoulli distribution $(1, \mu)$ with mean $\mu$.

2. A linear multilevel regression part with a latent (unobserved) predictor $\eta_{ji}$ of the binary outcome $x_{ji}$: $\eta_{ji} = \beta_0 + u_{0j}$, where $u_{0j}$ is a normally distributed random effect $u_{0j} \sim N(0, \sigma^2_{u0})$ with the variance $\sigma^2_{u0}$,

3. A link function connects the expected value of the dependent variable x with the latent predictor $\eta$, which is here the logit function: $\eta = logit(\mu) = \log(\mu/(1-\mu))$. Probabilities which range between 0 and 1 are transformed by the logit link function to logits, which continuously vary between $-\infty$ and $+\infty$ with a variance of $\pi^2/3 = 3.29$.

The multilevel logistic model for the observed proportions $p_j$ of papers which belong to the class 10% publications can be formulated as follows (Snijders and Bosker, 2004)

$$p_j = logistic(\beta_0 + u_{0j}) \qquad u_{0j} \sim N(0, \sigma^2_{u0}), \qquad (1)$$

where "logistic" means the logistic transformation of $p_j$ (logistic(x) = $e^x/(1+e^x)$), which is the inverse logit link function. There is a so called intra-class correlation between papers within institutions with $\rho = \sigma^2_{u0}/(3.29 + \sigma^2_{u0})$ which reflects the homogeneity of papers within an institution. The Wald test allows to test whether $\sigma^2_{u0}$ deviates from 0 (the null hypothesis). If the Wald test is statistically significant, the institutions systematically vary with respect to their number of class 10% papers. Then, a ranking or comparison of institutions is reasonable. Covariates can be included in the model in order to control, for instance, for socio-economic differences between countries (Bornmann et al., in press-c).

Most importantly, the multilevel model allows the calculation of so-called Empirical Bayes (EB) or shrinkage estimates which are more precise than their empirical counterparts, the raw probabilities. The following information is considered in the calculation of EB: First, if there is no further information for an institution, the mean value (i.e. mean probability) of class 10% papers across all institutions is the best estimate. Second, the more reliable the information for an institution (i.e. the greater the variance between institutions $\sigma^2_u$ and the higher the total number of papers for the institution under consideration), the more the raw probability of class 10% papers is the best estimate for this institution. The EB, therefore, vary between the mean value of class 10% papers across all institutions and the raw probability of class 10% papers for a certain institution. If the sample size (number of papers) for an institution is low, the EB is shrunken towards the mean value. The estimated standard errors take design effects into account (SAS Institute Inc., 2008). They are different from the corresponding standard errors and confidence intervals obtained in a sampling procedure which does not consider any clusters, especially where the sample size of level-2 units is small.

The EB and the confidence intervals can be transformed back to probabilities to facilitate the interpretation of the results. The multiplication of standard errors by 1.39 instead of 1.96 results in so-called Goldstein-adjusted confidence intervals (Goldstein and Healy, 1994) with the property that if the confidence intervals of two institutions do not overlap, they differ statistically significantly ($\alpha = 5\%$) in their estimates (i.e. class 10% papers' probabilities). If the 95%- confidence interval does not include the mean proportion of class 10% papers across all institutions, the authors located at this institution have published a statistically significantly greater or smaller number of class 10% papers than the average across all institutions. In case of Goldstein-adjusted confidence interval this test can only be done on the 16.3% probability level, rather than on the usual 5% level.

The power of statistical tests is defined as the probability of rejecting the null hypothesis in the case that is actually true. Based on simulation studies, Moineddin et al. (2007) recommended for a multilevel logistic regression at least 100 groups with a group size of 50 for an acceptable power. Except for the subject areas Psychology (n=59) as well as Pharmacology, Toxicology and Pharmaceutics (n=86) all subject areas in this study significantly exceeds this threshold for the number of groups (here: institutions). With respect to group size (here: publications) all subject areas exceeds the threshold of 50. We abstained from performing a power analysis here, because we reanalyzed with the publication and citation data observed data. Retrospective or post hoc power analyses do not provide useful and valid estimators of the true power (Levine and Ensom, 2001, Yuan and Maxwell, 2005). Simulation studies have shown that "a low power does not always indicate that the test is unpowered" (Sun et al., 2011, p. 81). Against this backdrop, we followed Levine and Ensom (2001) who pointed out that "confidence intervals better inform readers about the possibility of inadequate sample size than do post hoc power calculations" (p. 405).

The analyses for this study were calculated using the proc glimmix procedure implemented in the SAS statistical software (SAS Institute Inc., 2008).

### 2.4  Programming of the visualization

We have tried "to make a visualization that is attractive and informative, and yet conveys its own contingency and limitations" (Spiegelhalter et al., 2011, p. 1400). The following data and results of 17 multi-level regression models (one model for each subject area) have been used as an input for the visualization: (1) Number of papers published by authors located in an institution; (2) number of papers for an institution belonging to the class 10% papers in a subject area; (3) the "true" proportion of class 10% papers for an institution as the result of the multi-level model; (4) the proportion's confidence interval (lower and upper limits); and (5) the information whether an institution's proportion differ statistically significantly from the mean over all institutions in a subject area (the expected value). To rank the institutions within a subject area, the logarithmized quotient of the "true" proportion of class 10% papers for an institution and the "true" proportion across all institutions was calculated for each institution. The rationale for applying a logarithm is to provide comparable scales for values above and below the expected value; in other words, on our rank scale, an institution producing twice as many papers as expected is as far from the point for the expected value as an institution producing half as many as expected.

The maps used in the visualization are custom styled map tiles generated with TileMill (http://mapbox.com/tilemill/) based on Open Street Map (http://openstreetmap.org) data. For developing the data overlays, we used the polymaps library (http://polymaps.org/); the dynamic tables were realized with the help of the DataTables jquery plugin (http://datatables.net/).

## 3    Results

Figure 1 shows a screen shot of the web application visualizing the results of multi-level analyses for 17 different subject areas. The URL of the web application is as follows: http://www.excellencemapping.net. The web application is password-protected: The application can be used for research purposes only. The password can be received from the authors of this paper.

For a selected subject category (e.g. Physics and Astronomy), the map on the left-hand side of the screen shows a circle for each institution with a paper output greater than or equal to 500. Users can move the map to different regions by using the mouse (click and drag) and zoom in (or out) by using the mouse wheel. Country labels and map details appear only at zoom levels of a certain depth, primarily in order to facilitate perception of the data markers. Both moving and zooming can also be done by using the control buttons at the top left of the screen. The circle area for each institution on the map is proportional to the number of published papers in the respective subject area. For example, the Centre National de la Recherche Scientifique (CNRS) has the largest circle (in Europe) on the Physics and Astronomy map, highlighting the high output of papers in this subject area (see Figure 2). The circle colour indicates the proportion of class 10% papers for the respective institution using a diverging colour scale, from blue through grey to red (without any reference to statistical testing): If the proportion of class 10% papers for an institution is greater than the mean (expected) value across all institutions, its circle has a blue tint. Circles with red colours mark institutions with proportions of class 10% papers lower than the mean. Grey circles indicate a value close to the expected value.

On the right-hand side of the web application, all those institutions are listed which are considered in the multi-level model for a subject area (section "Institutional scores"). For each institution the name, the respective country, the number of all the papers published ("Papers"), and the EBs with confidence intervals are visualized ("Probability of excellent papers"). The greater the confidence interval, the more unreliable the probability for an institution is. If the confidence interval does not overlap with the mean proportion of class 10% papers across all institutions (the mean is visualized by the short line in the middle of "Probability of excellent papers"), the authors located at this institution have published a statistically significantly higher (or lower) number of class 10% papers than the average across all the institutions ($\alpha = 0.165$). The institutions in the list can be sorted (in decreasing or increasing order in case of numbers) by clicking on the relevant heading. Thus, the top or worst performers in a field can be identified by clicking on "Probability of excellent papers." Institutions with high productivity in terms of paper numbers appear at the top of the list (or at the end) by clicking on "Papers." In Biochemistry, Genetics and Molecular Biology, for example, the institution with the highest productivity between 2005 and 2009 is the CNRS; in terms of probabilities of class 10% papers, the best-performing institution is the Broad Institute of MIT and Harvard (see Figure 3). To reduce the set of visualized institutions in a field to only those which differs statistically significantly in their performance from the mean value, the corresponding tick mark can be set by the user.

Using the search field at the top right, the user can find a specific institution. By clicking on a heading (e.g. "Papers") the reduced list is ordered accordingly. To identify the institutions for a specific country, click on "Country". Then the institutions are first sorted by country and second by the probability of excellent papers (in increasing or decreasing order). "Your selection" is intended to be the section for the user to compare institutions of interest directly. If the confidence intervals of two institutions do not overlap, they differ statistically significantly on the 5% level in the probability of class 10% papers' output. For example, in Physics and Astronomy Stanford University and the Fraunhofer Gesellschaft are visualized without overlap (see Figure 4). The selected institutions in "Your selection" can be sorted by each heading in different orders. These institutions are also marked on the map with a black border. Thus, both institutional lists and institutional maps are linked by the section "Your selection". For the comparison of different institutions, it is not only possible to select them in the list but also on the map with a mouse click. A new comparison of institutions can be started by clicking on "Clear selection".

If the user has selected some institutions or has sorted them in a certain order, the selection and sort order are retained if the subject area is changed. This feature makes it possible to compare the results for certain institutions across different subject areas directly.

## 4    Discussion

The web application presented in this paper allows for an analysis which reveals centres of excellence in different fields worldwide using publication and citation data. Similar to the Leiden Ranking, only specific aspects of institutional performance are taken into account and other aspects such as teaching performance or the societal impact of research (Bornmann, 2012, Bornmann, 2013) are not considered (see Waltman et al., 2012). Based on data gathered from Scopus, field-specific excellence can be identified in institutions where highly-cited papers have been published frequently. The web application combines both, a list of institutions ordered by different indicator values and a map with circles visualizing indicator values for geocoded institutions. Compared to the mapping and ranking approaches introduced hitherto, our underlying statistic (multi-level models) are analytically oriented by allowing (1) the estimation of statistically more appropriate values for the number of excellent papers for an institution than the observed values; (2) the calculation of confidence intervals as reliability measures for the institutional citation impact; (3) the comparison of a single institution with an "average" institution in a subject area and (4) the direct comparison of at least two institutions. With these features, our approach can not only identify the top performers in (excellent) output but the "true jewels" in different disciplines. These are institutions having published statistically significantly more class 10% papers than an "average" institution in a subject area. Against the backdrop of these advantages, our web application can be used by (1) scientists for exploring excellence centres and regions worldwide in a specific subject area, (2) students and parents to get helpful hints about the comparative merits of different universities for making choices, and (3) governments and policymakers to compare the performance of institutions within a specific country to those outside (see Hazelkorn, 2011).

Despite the advantages of the web application to map excellence in science, we recognize the limitations inherent in bibliometric data. (1) Papers are only one among several types of scientific activities. Research is a multi-dimensional endeavour which cannot be captured with only a single indicator. (2) It is not guaranteed that the addresses listed on the publication reflect the locations where the reported research was conducted. There might be several addresses on a publication but the research was mainly conducted at one location. (3) No standard technique exists for the subject classification of articles (Bornmann and Daniel, 2008, Bornmann et al., 2008, Leydesdorff and Rafols, 2009). For the web application, we have used the standard technique based on journal classification schemes. Although this approach has been frequently criticized and other solutions have been proposed (e.g. the categorization of papers based on index terms from field-specific databases), it has not been possible to establish any other proposals as a (new) standard for interdisciplinary studies up to now.

Besides the limitations inherent in bibliometric data, there are several problems inherent in mapping approaches such as that proposed here. A detailed list of these problems is published in Bornmann et al. (2011a). The user should always be aware of these limitations when the web application is used. For example, there may be circles for institutions on the maps that are not in the right position. In the various routines, we tried to avoid these misallocations, but they could not be completely resolved. The misallocations do have different sources: address errors in the Scopus data or erroneous coordinates provided by the geocoding process. Furthermore, high numbers of excellent papers visualized on the map for a single institution might be due to the following two effects: (a) Many scientists located in that institution produced at least one excellent paper each or (b) only a few scientists located in this institution produced many influential papers. Assuming institutions as units of analysis, one is not able to distinguish between these two interpretations.

The web application described here allows future developments in several directions. (1) Further data sets can be uploaded, e.g. for the visualization of patent data (Leydesdorff and Bornmann, 2012a). (2) With multi-level models it is possible to consider data for more than one subject area. (3) This data can be used to categorize institutions in different groups such as universal performers which are successful in all subject areas and specific performers which are successful only in some areas (Bornmann et al., in press-b). Covariates might be included into the regression models to control class 10% probabilities for certain factors (e.g., institutional size, economic force) (Bornmann et al., in press-c). (4) The evaluation of spatial performance could be extended by network approaches proposed by Hennemann (2012) "in which co-authorships of scientific papers represent a definable relationship between knowledge-producing players in a system" (p. 2402) and by Calero Valdez et al. (2012) who proposed a performance metric of interdisciplinarity. According to Mazloumian et al. (2013)

"given the large relevance of network theory in many scientific areas, we believe that classical, *node-based indices* must be complemented by *network-based indices*."

Table 1. Number of universities and research-focused institutions included in the statistical analyses for 17 different subject areas. The mean percentage of highly-cited papers is the mean over the percentages of class 10% papers for the institutions within one subject area.

| Subject area | Number of universities or research-focused institutions | Mean percentage of highly-cited papers (class 10% papers) |
| --- | --- | --- |
| Agricultural and Biological Science | 504 | 0.15 |
| Biochemistry, Genetics and Molecular Biology | 746 | 0.13 |
| Chemical Engineering | 148 | 0.14 |
| Chemistry | 496 | 0.13 |
| Computer Science | 350 | 0.14 |
| Earth and Planetary Sciences | 318 | 0.17 |
| Engineering | 594 | 0.14 |
| Environmental Science | 215 | 0.17 |
| Immunology and Microbiology | 204 | 0.16 |
| Materials Science | 367 | 0.14 |
| Mathematics | 362 | 0.14 |
| Medicine | 1175 | 0.17 |
| Neuroscience | 108 | 0.17 |
| Pharmacology, Toxicology and Pharmaceutics | 86 | 0.17 |
| Physics and Astronomy | 650 | 0.14 |
| Psychology | 59 | 0.20 |
| Social Sciences | 166 | 0.19 |

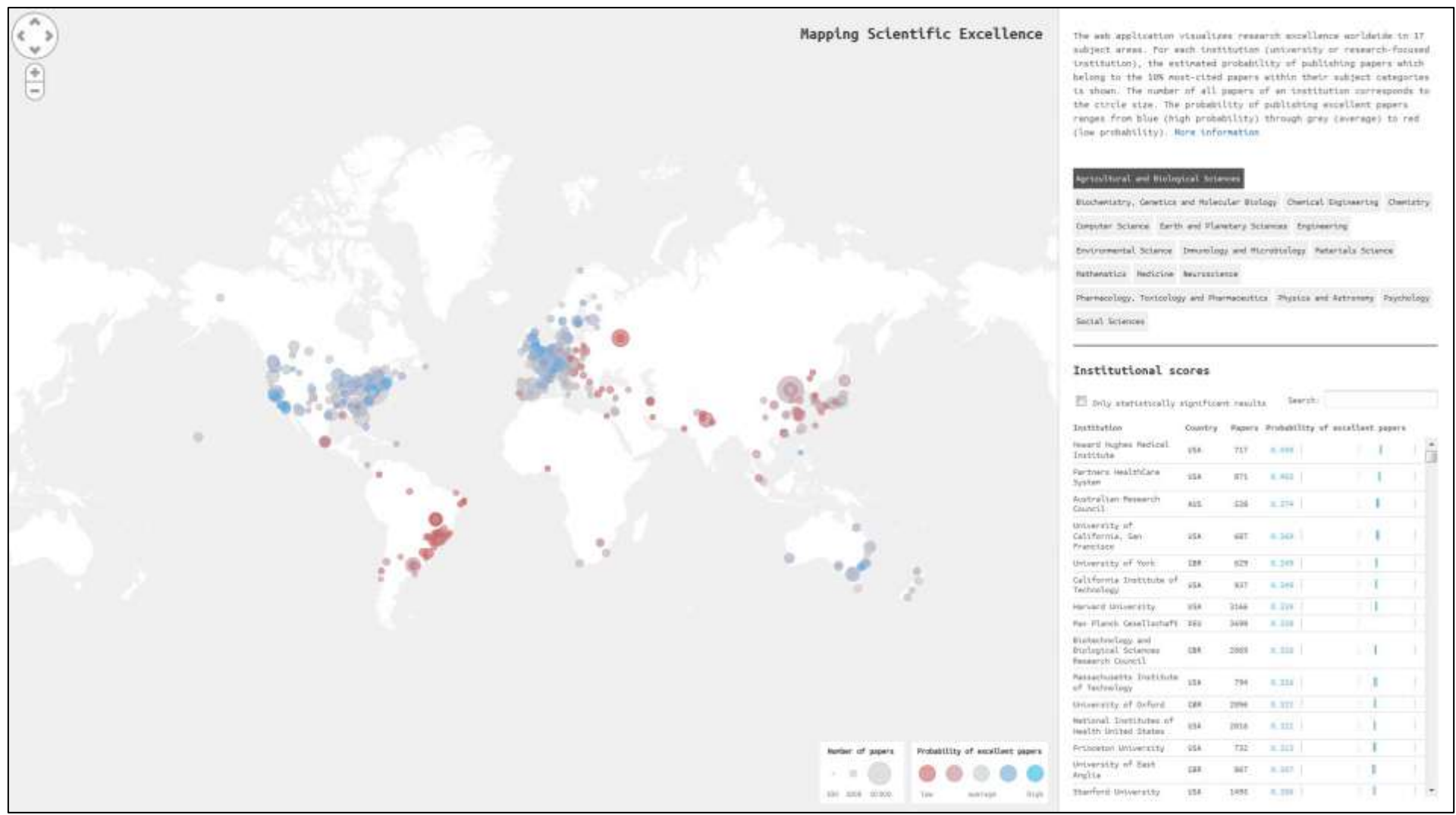

Figure 1. Screen shot of the web application

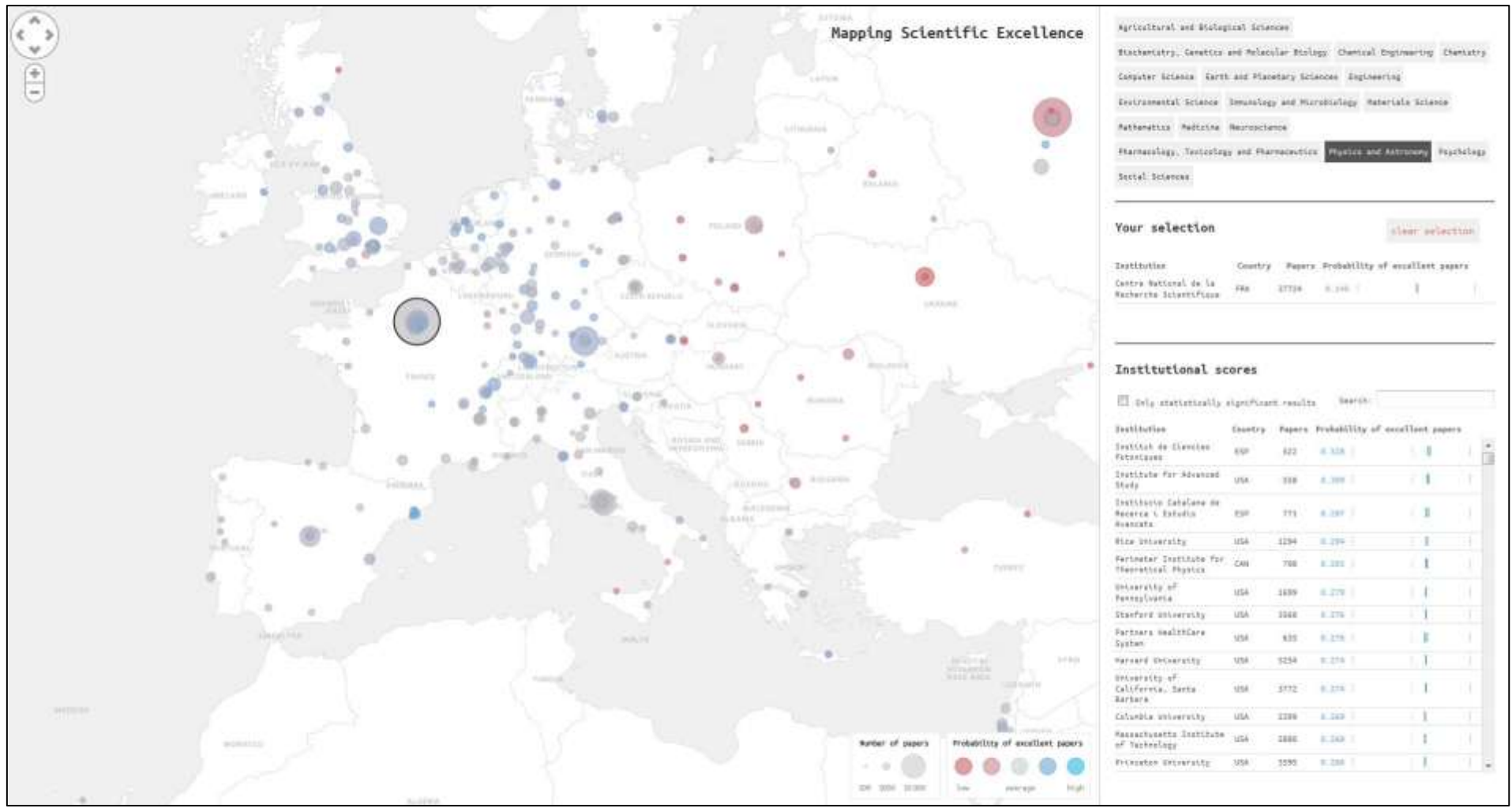

Figure 2. Physics and Astronomy map where the Centre National de la Recherche Scientifique is selected.

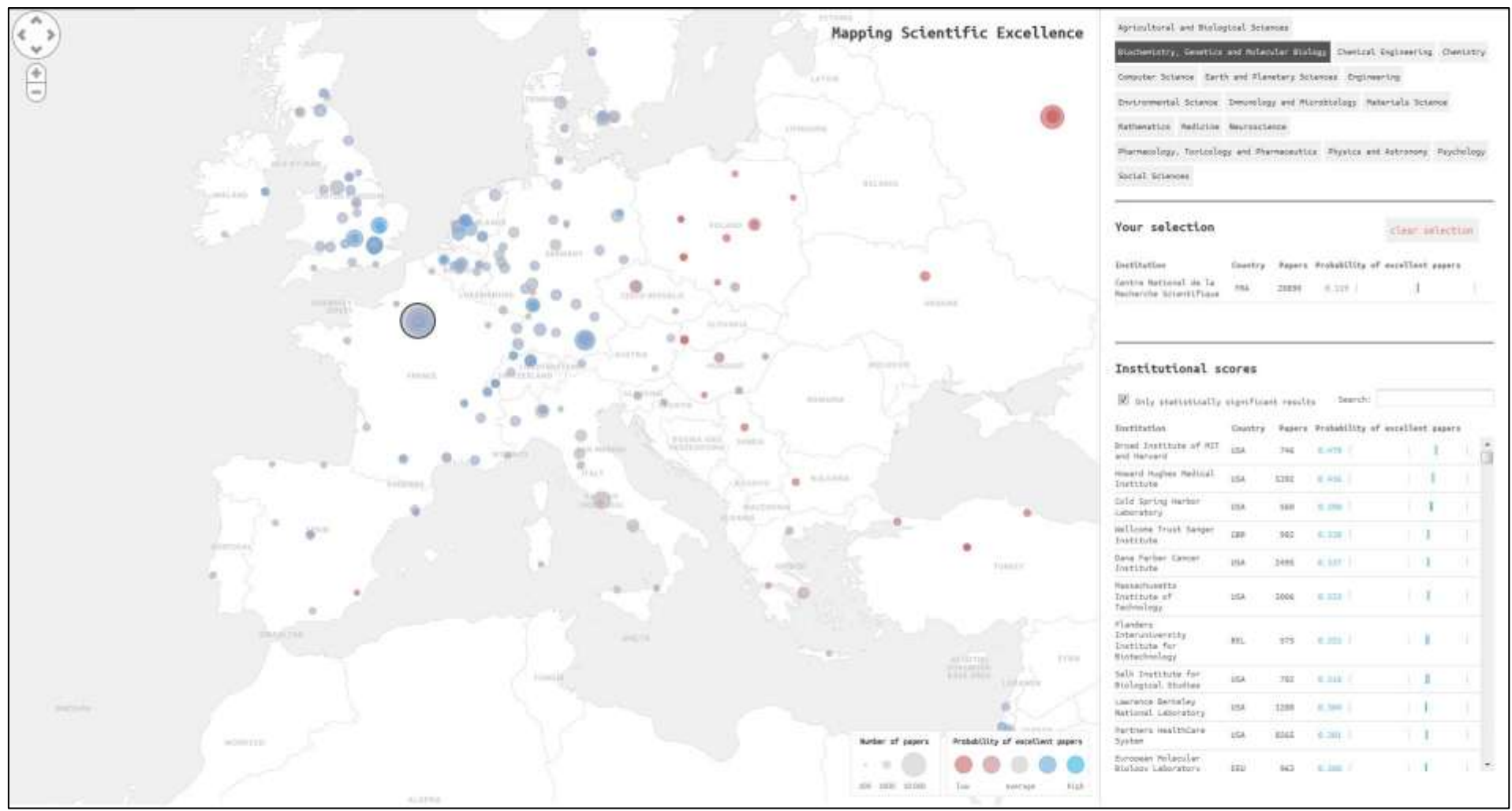

Figure 3. Biochemistry, Genetics and Molecular Biology map where the Centre National de la Recherche Scientifique is selected.

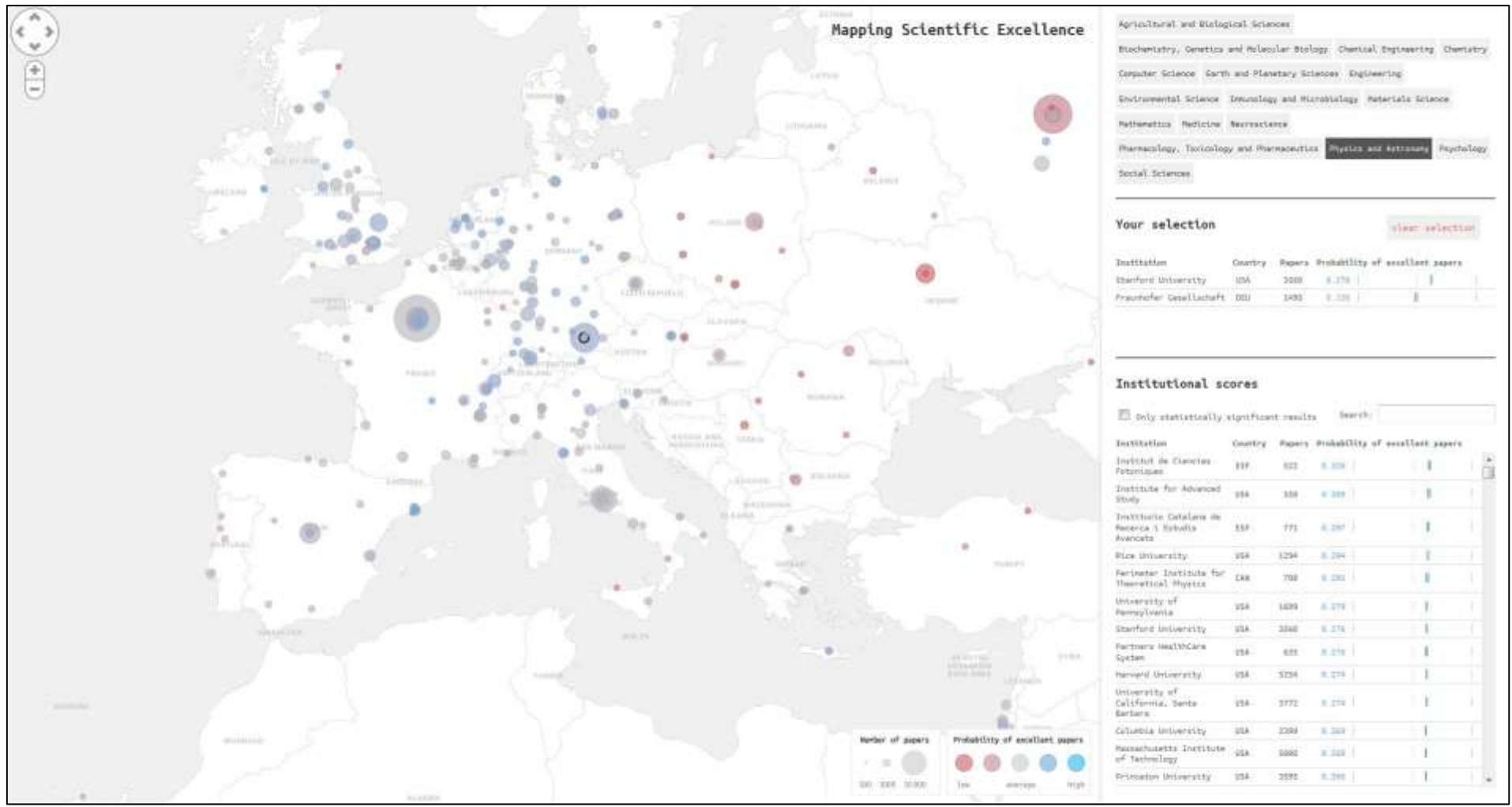

Figure 4. Physics and Astronomy map where the Stanford University and Fraunhofer Gesellschaft are selected.